\documentclass[superscriptaddress,aps,prl, amsmath,amssymb,reprint,showkeys,showpacs]{revtex4-2}

\usepackage{graphicx}
\usepackage{dcolumn}
\usepackage{bm}
\usepackage{lipsum}
\usepackage{physics}
\usepackage{xcolor}

\usepackage{hyperref}
\usepackage{here}

\usepackage[mathlines]{lineno}

\begin{document}

\preprint{APS/123-QED}

\title{Post-Selection Free Generation of Multi-Photon Added Coherent States}

\author{M. Uria}
\email{maruria@udec.cl}
\affiliation{Departamento de F\'isica, Facultad de Ciencias F\'isicas y Matem\'aticas, Universidad de Concepci\'on, Concepci\'on, Chile}
\affiliation{ANID - Millenium Science Iniciative Program - Millenium Institute for Research in Optics}

\author{R. Guti\'{e}rrez-J\'{a}uregui}
\affiliation{Departamento de F\'isica Cu\'antica y Fot\'onica, Instituto de F\'isica,Universidad Nacional Aut\'onoma de M\'exico, Ciudad de M\'exico, 04510, M\'exico}

\author{C. Hermann-Avigliano}
\affiliation{ANID - Millenium Science Iniciative Program - Millenium Institute for Research in Optics}
\affiliation{Departamento de F\'{\i}sica,  Facultad
de Ciencias F\'isicas y Matem\'aticas, Universidad de Chile,
Santiago, Chile}

\author{P. Solano}
\email{psolano@udec.cl}
\affiliation{Departamento de F\'isica, Facultad de Ciencias F\'isicas y Matem\'aticas, Universidad de Concepci\'on, Concepci\'on, Chile}

\date{\today}

\begin{abstract}

Non-Gaussian quantum states are essential resources for continuous-variable quantum information processing and for metrology. Among these, multi-photon added coherent states bridge classical and non-classical behaviors; however, their generation typically relies on small photon numbers and probabilistic heralding schemes. Here, we propose a protocol for the post-selection free generation of high fidelity  multi-photon added coherent states using the photon blockade effect in a driven Kerr nonlinear resonator, where such states emerge naturally during the dynamics. We demonstrate that high-fidelity states can be prepared by optimizing the external drive power and the interaction time. Furthermore, we show that the protocol is robust under realistic experimental conditions, achieving fidelities of $\approx 99\%$ with current state-of-the-art parameters. Our results unlock a deterministic route to complex non-classical states using well-established quantum optical platforms.

\end{abstract}


\maketitle

{\it Introduction.---} The field of quantum optics has developed around the inherent statistical nature of the electromagnetic field,  exploring different states of light whose fluctuations can be controlled to perform particular tasks \cite{Hahn_2022,Nielsen2007,Sperling2014,Pizzimenti2021}. Two contrasting states of light are coherent and number states, which are quintessential examples of dissimilar statistics in quantum optics. Coherent states are characterized by minimizing fluctuations in the field quadratures,  which have prompted them to be considered as nearly classical states~\cite{Glauber1963}. When used to illuminate matter, these states enable the buildup of atomic coherences, opening a window into the quantum world.  In comparison, number states display a definite photon number, which has been leveraged to generate entanglement with matter~\cite{Haroche_2006} and explore the role of measurements in quantum systems~\cite{Solano_2005,Laiho_2009}. The transition between both limits can occur continuously, as encapsulated by photon added coherent states (PACS) \cite{Agarwal_1991}. These states, obtained by repeatedly adding single photons to a coherent state, bridge the operational regimes of continuous and discrete variable quantum optics.

Owing to their unique intermediate character, PACS are highly non-classical. They display sub-Poissonian statistics, squeezing, and Wigner function negativity, making them important candidates for quantum communication, computing, and metrology \cite{Braunstein2005,Pinheiro_2012,Chatterjee_2021,Kudra_2025}.  However, PACS implementations have been limited to small numbers of added photons and post-selective methods, with proposals obtained for single photons \cite{Zavatta2004, Barbieri2010, Kumar2013} and state-of-the-art experiments for up to three photons  \cite{Fadrny_2024}. Recent works have shown strategies to produce arbitrarily large photon-number states without post-selection \cite{Uria2020,Kudra_2022,Heeres_2015} by controlling the timing and strength of light-matter interactions. Although such protocols are not directly applicable to the problem of generating multi-photon PACS, they point towards the pathway explored here. 

\begin{figure}
    \hspace{-0.5cm}\centering
    \includegraphics[width=0.9\linewidth]{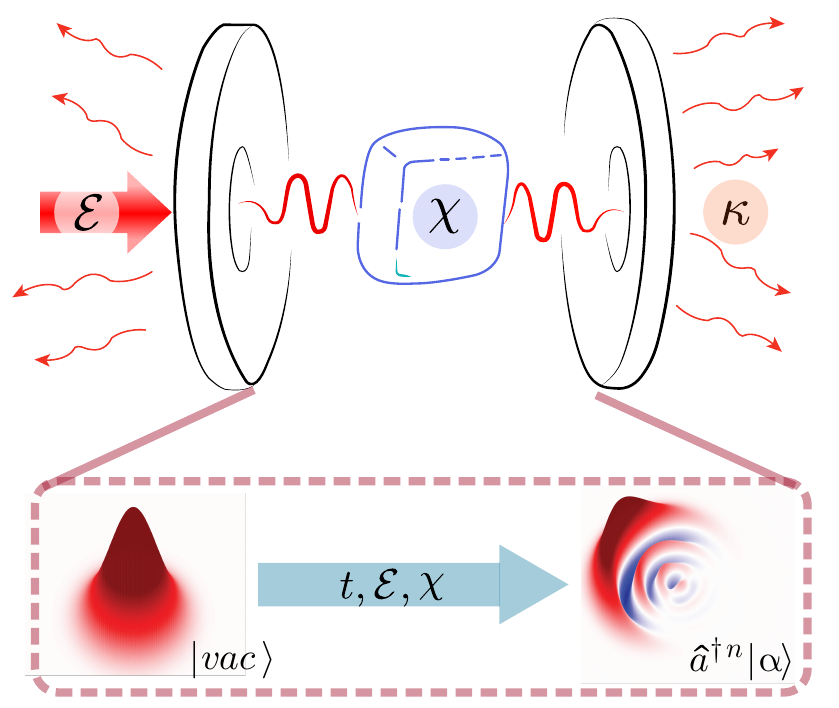}
    \caption{Schematic of the protocol for high fidelity generation of PACS. A coherent drive of amplitude $\mathcal{E}$ excites a single-mode cavity with Kerr nonlinearity $\chi$ and decay rate $\kappa$. Starting from the vacuum state $\ket{vac}$, the system reaches a state resembling a displaced multi-photon added coherent state $\hat{a}^{\dagger n}\ket{\alpha}$ at a time $t$. The lower panel illustrates the transformation in phase space, highlighting the emergence of non-classical features from an initially Gaussian state.}
    \label{fig:fig1}
\end{figure}

In this Letter, we present a protocol for generating high-fidelity multi-photon PACS without post-selection. The proposed platform consists of a cavity with Kerr-type nonlinearity driven by an external coherent electromagnetic field (as shown in Fig. \ref{fig:fig1}). This setup contains the ingredients to form photon added states, with the nonlinearity organizing the system around states of well-defined photon numbers, while the external drive organizes them along a defined phase. We show how displaced PACS are naturally formed by the interplay of these parameters. Such a small, but reversible, displacement veils this emerging state in standard quantum optics treatments of such a well-studied platform. To address this reversible change, we propose a witness that identifies the added photons regardless of any coherent displacement in the quadrature space. By monitoring the dynamics of this witness, we identify the evolution time at which a displaced PACS emerges and then determine the optimal coherent displacement required to finalize with a $N$-photon PACS. Our simulations find fidelities greater than $99\%$ for PACS with at least 10 added photons, limited by our numerical exploration. We then proceed to study the effects of dissipation to show the feasibility of the protocol using parameters from recent circuit-QED experiments \cite{Iyama_2024}. We conclude by discussing the results of the protocol and providing a glimpse of some open questions.

{\it PACS definition.---} Coherent states of light $|\alpha\rangle$ are quantum states characterized by intensity correlations at all orders and a Poissonian photon-number distribution, which makes them nearly classical states \cite{Glauber1963}. In contrast, photon-number states $|n\rangle$ are the archetypic quantum states with particle-like intensity correlations without classical analogs, such as antibunching \cite{Agarwal2012}. Photon-added coherent states (PACS) smoothly connect both contrasting examples \cite{Agarwal_1991}. They are obtained by applying $n$ times the creation operator, $\hat{a}^\dagger$, to a coherent state as
\begin{equation}
    \ket{\alpha,n}=\frac{\hat{a}^{\dagger n} \ket{\alpha}}{\sqrt{N_n}},
\end{equation}
where $n$ is an integer that determines the number of added photons and $N_n=n!L_n(-\abs{\alpha}^2)$ is a normalization constant, with $L_n(x)$ being the Laguerre polynomial of order $n$ \cite{Cahill_1969}. In the limit $\alpha \rightarrow 0$, PACS tends to a photon-number state of $n$ photons, while in the limit of large $\alpha$,  adding photons to a coherent state has negligible effects on its statistics, creating a crossover between non-classical to classical states as a function of $\alpha$. 

PACS can also be written in the photon-number basis as \cite{Agarwal_1991}
\begin{equation}
\begin{split}
    \ket{\alpha,n}&=\frac{\hat{D}(\alpha)}{\sqrt{N_n}}\sum_{s=0}^n \binom{n}{s} \sqrt{s!}(\alpha^*)^{n-s} \ket{s}\\
    &=\hat{D}(\alpha)\ket{\Gamma_{\alpha,n}},
\end{split}
    \label{eq:PACS}
\end{equation}
where $\hat{D}(\alpha)$ is the coherent displacement operator. Because the displacement operator is a linear transformation, $\ket{\Gamma_{\alpha,n}}$ preserves the essential features of the state, such as its non-Gaussian character. This notation allows us to relate PACS to a truncated Hilbert space \cite{Sivakumar2014}, with an upper bound $n$, and suggests that systems with strong photon blockade \cite{Imamo_1997,Hoffman2011} could lead to PACS generation under right conditions.

{\it Theoretical model.---} We consider a driven single-mode cavity containing a Kerr nonlinear medium in order to realize photon blockade.  The evolution of the electromagnetic field inside the cavity is well described by the Drummond and Walls model \cite{Drummond_1980}, given by the Hamiltonian
\begin{equation}
\mathcal{H}=\hbar \omega_0 \hat{a}^{\dagger}\hat{a} +\hbar \mathcal{E}\left( \hat{a}\ e^{i\omega_d t} + \hat{a}^{\dagger} e^{-i\omega_d t} \right) + \hbar\chi \hat{a}^{\dagger 2}\hat{a}^2,
\end{equation}
where $\omega_0$ ($\omega_d$) is the frequency of the cavity (driving) field, $\mathcal{E}$  the driving amplitude, and $\chi$ the nonlinearity strength. Working within a rotating frame with respect to the drive frequency $\omega_d$, we obtain
\begin{equation}
    \mathcal{H}_R = \hbar  \delta\,\hat{a}^{\dagger}\hat{a} +\hbar \mathcal{E} (\hat{a} + \hat{a}^{\dagger})+\hbar\chi\, \hat{a}^{\dagger2}\hat{a}^2,
    \label{eq:H}
\end{equation}
with detuning $\delta=\omega_0-\omega_d$. 

This Hamiltonian captures the competition between two mechanisms: the coherent drive increases the number of excitations along a defined phase in the quadrature space within the cavity, while the nonlinearity shifts the cavity out of resonance by activating upper number states, suppressing additional excitations. This competition between the drive and Kerr nonlinearity leads to a photon blockade \cite{Miranowicz2013}, where the creation of additional photons in the cavity mode is inhibited \cite{Imamo_1997, Anikin_2021}.

\begin{figure}
    \centering
    \includegraphics[width=1.0\linewidth]{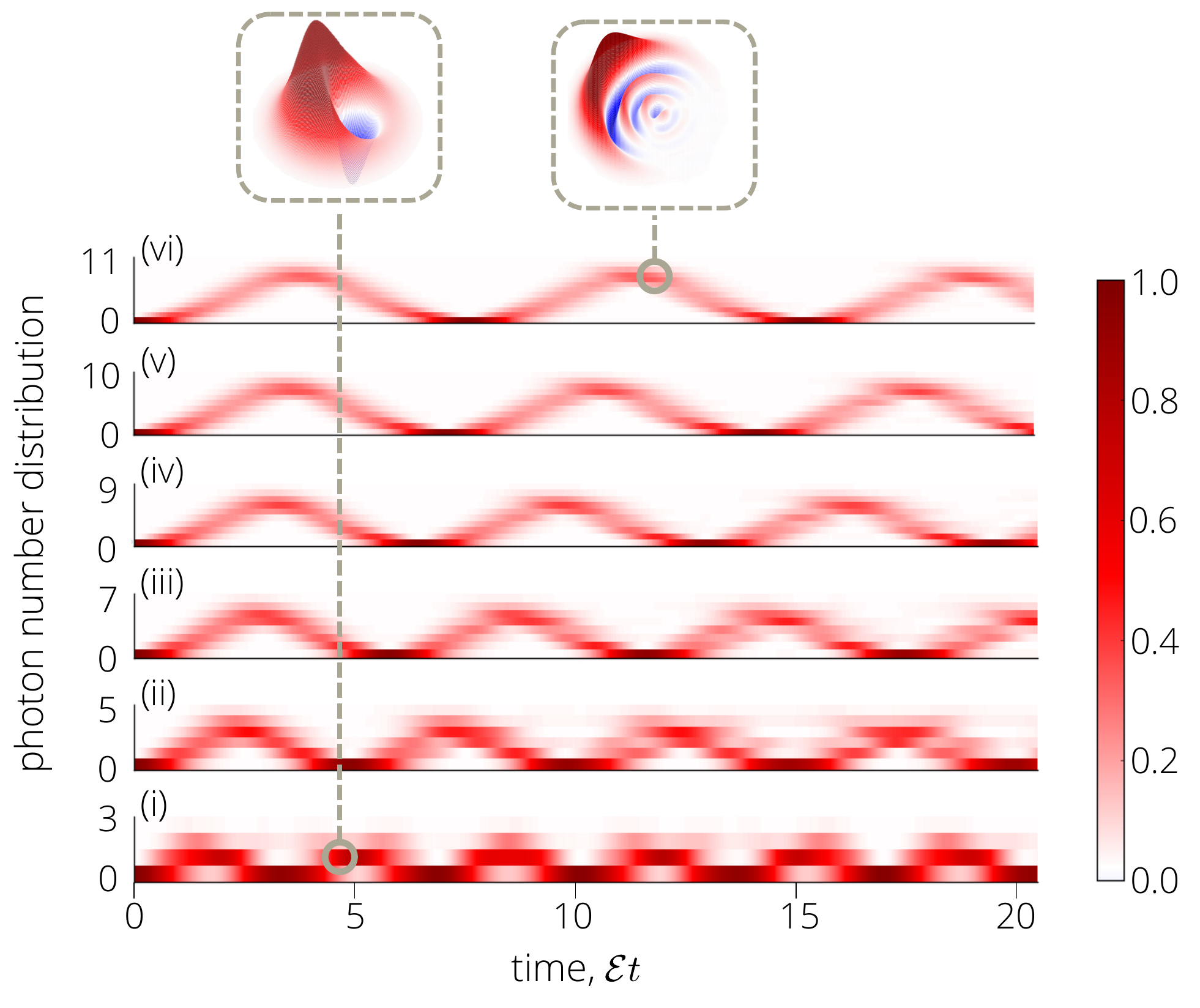}
    \caption{Photon number distribution as a function of evolution time for different driving values ($\mathcal{E}/\chi=1.5,4.5,7.25,10,13,16$) and fixed $\delta=2\chi$. The roman numerals represent the number of added photons for the best PACS generated in each driving regime. The insets on the top represent the Wigner function of the best PACS found in the left ($n=1$) and right ($n=6$) cases.}
    \label{fig:dist}
\end{figure}

We consider the system initialized in the vacuum state $\ket{\psi_{\text{sys}}(0)}=\ket{0}$ and out of resonance ($\omega_0\neq \omega_d$). Figure \ref{fig:dist} shows the time evolution of the photon-number distribution for different driving amplitudes and fixed detuning $\delta=2\chi$ (meaning that the field linear evolution frequency coincides with the smallest Kerr evolution frequency, see Supplementary Material \cite{SM}). The presence of an upper bound in the photon-number distribution evidences the effect of photon blockade. The insets in the upper part correspond to the Wigner function distributions of the states that closely resemble a displaced $\ket{\Gamma_{\alpha,n}}$ and produce PACS with high fidelity. These states emerge almost periodically, appearing at times determined by an interplay of drive amplitude and nonlinearity, suggesting the creation of dressed states connecting ground and fully blockaded states. They emerge almost when the photon number $\expval{\hat{n}}$ is maximum, or more precisely when the remnant photon number  (defined bellow) is maximized.

{\it Remnant photon number.---} As discussed above, the cavity field periodically evolves into a blockaded state that resembles $\hat{D}(\beta)\ket{\Gamma_{\alpha,n}}$. This state retains the fundamental non-classicality of a PACS [Eq. (\ref{eq:PACS})] but follows a complex trajectory in the phase space due to the nonlinear dynamics. The followed trajectory creates a moving-target, whose fidelity respect to a PACS is difficult to compute, since there is no analytical solution for $\beta(t)$.

To bypass this limitation, we define the \textit{remnant photon number} as a frame-independent witness of state formation, given by
\begin{equation}
n_{r}=\langle(\Delta \hat{x}_{\theta})^2\rangle+\langle(\Delta \hat{x}_{\theta+\pi/2})^2\rangle-1/2.
\label{eq:meas}
\end{equation}
Here, $\hat{x}_{\theta}=(\hat{a}e^{-i\theta}+\hat{a}^\dagger e^{i\theta})/2$  is an arbitrary quadrature of the field, and $\langle(\Delta \cdot )^2\rangle$ represents the variance. Crucially, $n_r$ measures the total fluctuations of the amplitude and depends only on the shape of the Wigner function, rendering it invariant under both rotation and coherent displacement \cite{Yadin_2018}. This quantity corresponds to subtracting a constant of 1/2 from the \textit{total noise} \cite{Hillery_1989}, but it also illustrates another physical aspect. Physically, $n_r$ represents the remnant photon number: the minimum mean photon number remaining in the field after an optimal displacement operation effectively centers the state at the origin of the phase space (see Supplementary Material \cite{SM}). For example, $n_r=0$ for any coherent state $|\alpha \rangle$ and $n_r=n$ for the Fock state $|n\rangle$. As such, $n_r$ can thus be regarded as a measure of non-classicality for pure states \cite{Hillery_1989}, related to its quantum Fisher information \cite{Tan_2019}. 

Consequently, we can use $n_r$ to detect the emergence of the Fock-like structure inherent to PACS, independent of the displacement of the state. By monitoring the evolution, we identify the interaction times where $n_r$ is maximized, signaling the formation of a PACS upon rectifying coherent displacement.

\begin{figure}
    \centering
\hspace{-0.5cm}\includegraphics[width=1.0\linewidth]{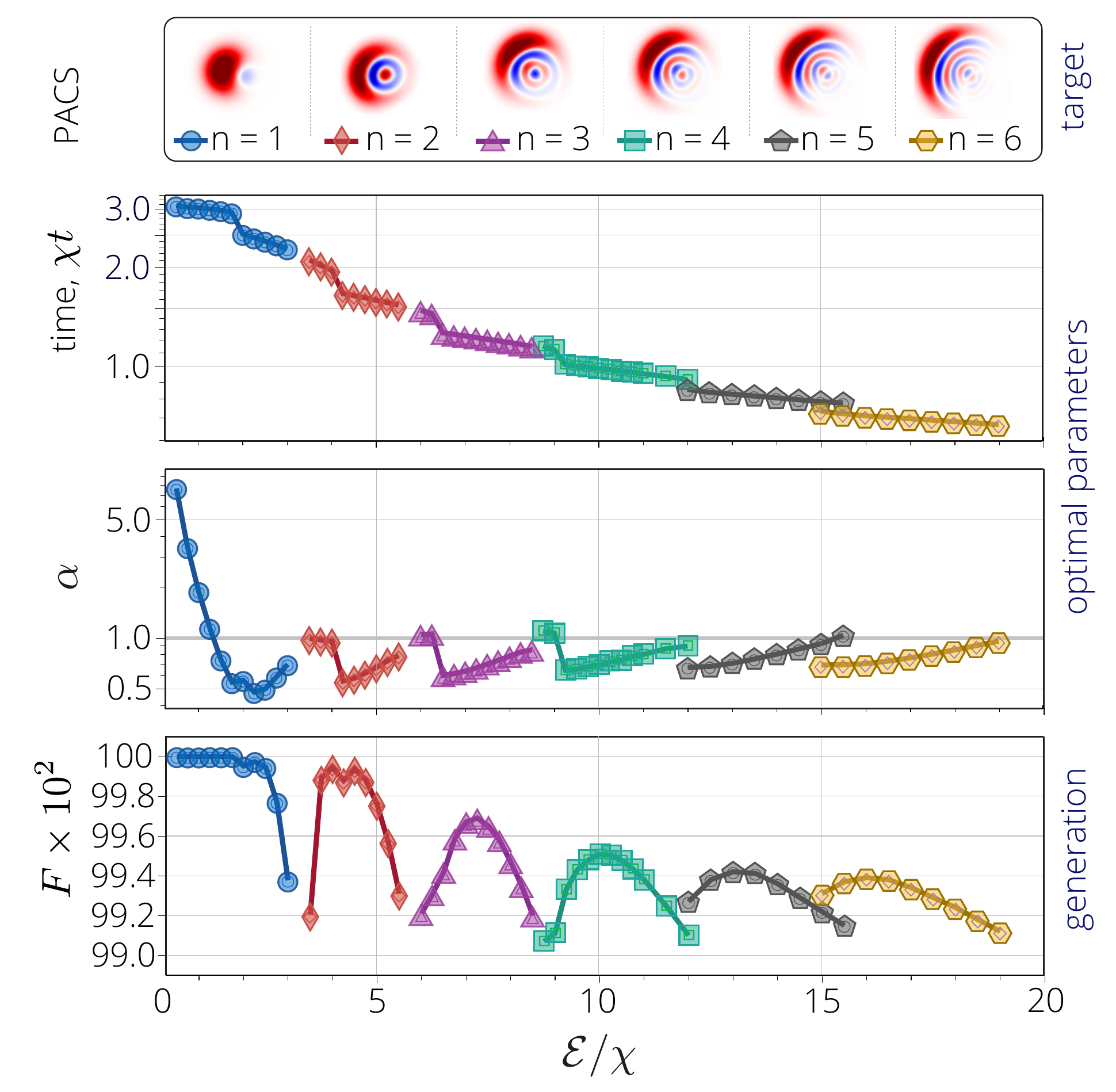}
    \caption{Fidelities and optimal parameters for PACS generation. The horizontal axis represents the drive amplitude $\mathcal{E}$ normalized by the nonlinear parameter $\chi$. From top to bottom: Wigner function representations, optimal evolution time, amplitude of the coherent component, and optimal fidelity of the obtained PACS. The shapes connected by colored curves represent the different number of photons added $n$ in the obtained PACS.  We considered a drive detuning $\delta=2\chi$ for all the calculations.}
    \label{fig:results2}
\end{figure}

{\it Parameters optimization and results.---} The cavity field state $\ket{\psi{(t)}}$ evolves almost periodically to a state close to displaced $\ket{\Gamma_{\alpha,n}}$. In order to obtain a PACS $\ket{\alpha,n}$, $\ket{\psi{(t)}}$ must be corrected by a coherent displacement with an unknown amplitude and phase. To find such a displacement, we first center $\ket{\psi(t)}$ in the phase space by applying a displacement $\hat{D}(-\gamma)$, with a measurable quantity $\gamma=\left<\hat{x}_{0}(t)\right>+i \left<\hat{x}_{\pi/2}(t)\right>$. The resulting state must then be displaced by $\hat{D}(\zeta)$ to approximate the PACS $\ket{\alpha,n}$. This means that we want to compute the fidelity of the state $\ket{\Psi(t)}=\hat{D}(\zeta )\hat{D}(-\gamma)\ket{\psi{(t)}}$ compared to $\ket{\alpha,n}$, defined as $F(t)=\abs{\braket{\alpha,n}{\Psi(t)}}^2$. 

We emphasize that neither the displacement $\zeta$ nor the coherent state amplitude $\alpha$ are known, and their values depend on the state evolution. Given a value of $\ket{\psi(t)}$, we can find the values of $\alpha$ and $\zeta$ that optimize the PACS generation by maximizing $F$. Without loss of generality, we define the complex phase of $\alpha$ such that it has the same orientation as the phase space representation of the obtained state $\ket{\Psi(t)}$, so that $\zeta$ and $\alpha$ share a common phase $\phi(t)$. Thus, the optimization reduces to the parameters $|\zeta|$, $|\alpha|$, and $\phi$.

To reduce the number of optimization parameters, we propose a heuristic model based on the evolution of the Wigner function of the field. We consider an ansatz for the complex phase $\phi(t)$, which is determined by the vector normal to the state trajectory in the phase space. There is an additional physical restriction: the number of photons should be the same for both $\ket{\Psi(t)}$ and $\ket{\alpha,n}$, allowing us to establish a relationship between the magnitudes of $\zeta$ and $\alpha$. This relationship is given by
\begin{equation}
\zeta(\alpha)=\left<\hat{x}_\phi\right>_{\alpha,n}\pm \sqrt{n_r(\ket{\psi(t)})-n_{r}(\ket{\alpha,n})},
\end{equation}
where $\left<\hat{x}_\phi\right>_{\alpha,n}$ is the quadrature expectation value of the PACS $\ket{\alpha,n}$ and $n_{r}(\ket{\alpha,n})=N_{n+1}/N_n-\left<\hat{x}_\phi\right>^2-1$ is the remnant photon number for the PACS, with $N_n$ being the normalization constant in Eq. (\ref{eq:PACS}). Finally, there is only one free parameter to optimize $|\alpha|$. 

The procedure described above is ad first heuristic approximation and may not yield the highest fidelity. Therefore, we refine the result by varying $\phi$ around the guessed value after optimizing for $|\alpha|$. Our simple optimization procedure already produces PACS with large added photon numbers and fidelities exceeding $99\%$.

Figure \ref{fig:results2} shows the fidelities and best parameters that optimize the procedure. These parameters were determined for a specific condition $\delta=2\chi$. Out of this condition, the fidelity decays below $99\%$. In all cases, the optimization is localized in times around the second maxima of the remnant number of photons $n_r(\ket{\psi})$. Figure~\ref{fig:results2} shows examples of the Wigner function representation of the generated states for increasing photon number, as well as the required evolution time, coherent amplitude $|\alpha|$, and fidelities for the generated PACS with up to 6 photons.

{\it Experimental feasibility.---} In a realistic scenario, dissipation must be considered. In the rotating frame, the photonic state $\rho$ of a lossy cavity evolves according to the master equation
\begin{equation}
    \begin{split}
        \frac{d\rho}{dt}&= \frac{1}{i\hbar}\left[\mathcal{H}_R,\rho\right] + \kappa(n_{\text{th}}+1)\mathcal{L}_{\hat{a}}[\rho] + \kappa n_{\text{th}}\mathcal{L}_{\hat{a}^{\dagger}}[\rho]  \\
    \end{split}
    \label{eq:dec}
\end{equation}
where $\mathcal{L}_{\xi}[\bullet] = 2\xi \bullet \xi^{\dagger} -  \bullet \xi^{\dagger}\xi - \xi^{\dagger}\xi \bullet $ is the Lindblad superoperator that describes incoherent processes. The parameter $\kappa$ characterizes the cavity loss rate, while $n_{\text{th}}$ is the mean thermal photon number, which accounts for finite-temperature effects.

\begin{figure}[ht]
    \centering
    \hspace{-0.1cm}\includegraphics[width=\columnwidth]{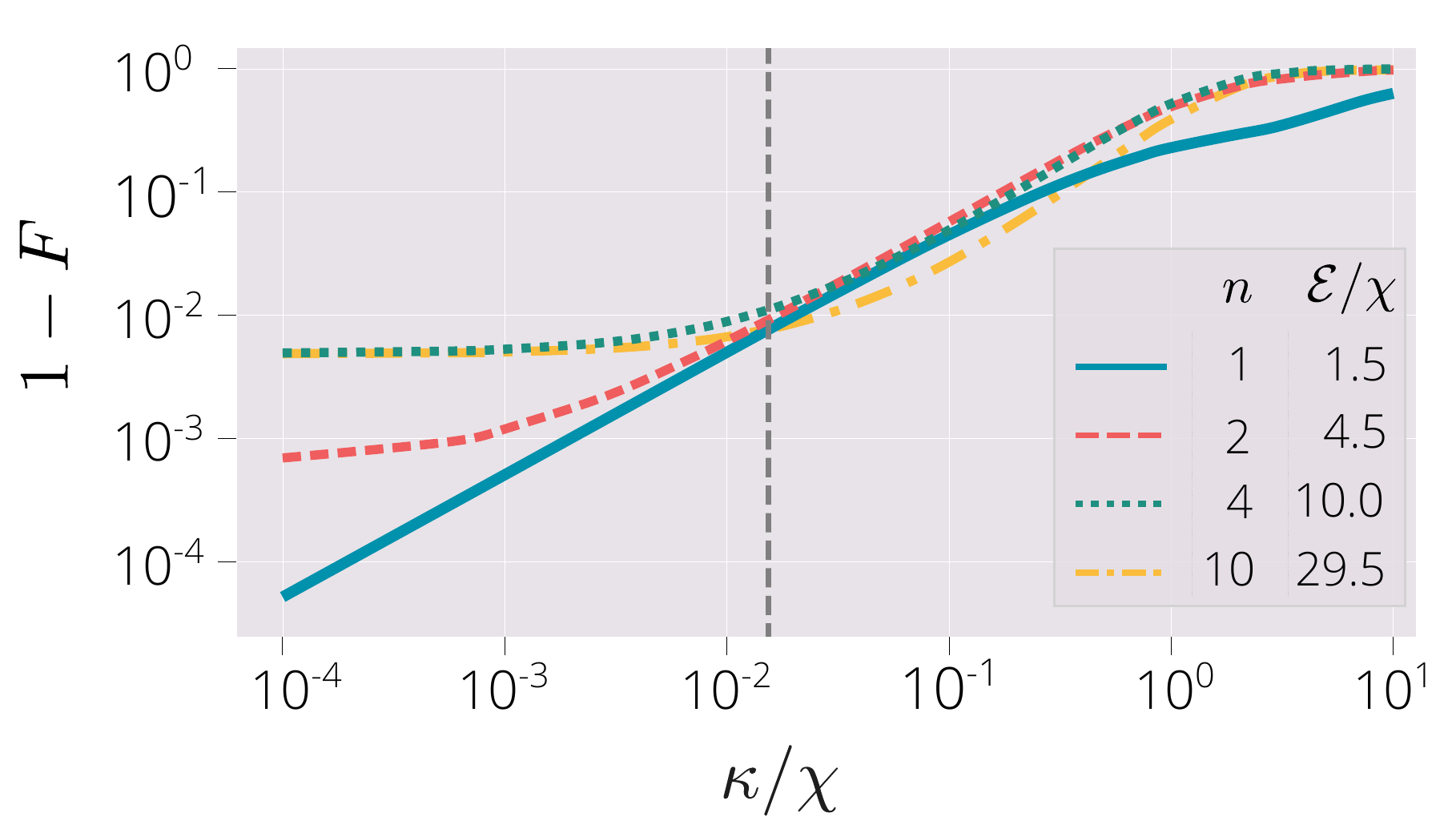}
    \caption{Infidelity for the optimal cases under decoherence for different driving amplitudes $\mathcal{E}$, with optimized numbers of added photons $n$. The vertical dashed gray line indicates the experimental state-of-the-art relation between the decoherence rate $\kappa$ and the nonlinearity $\chi$, as reported in Ref.~\cite{Iyama_2024}.}
    \label{fig:deco}
\end{figure}

Cavity losses degrade the purity of the evolving state. To assess the experimental feasibility of generating PACS, we numerically solve Eq. (\ref{eq:dec}), apply the necessary coherent displacement to generate a PACS, and compute its infidelity ($1-F$) from the target state. Figure \ref{fig:deco} shows the infidelity as a function of the cavity decay rate $\kappa$ for different driving amplitudes $\mathcal{E}$. A vertical gray line indicates the experimental parameters for state-of-the-art experimental realizations \cite{Iyama_2024}.  We conclude that generating useful PACS with fidelities around 99\%, is currently feasible.

{\it Analysis and discussion.---} The deterministic generation of PACS relies on the interplay between photon blockade and the phase-space dynamics. The detuned drive creates an effective rotation around a displaced origin in the phase space, while the Kerr nonlinearity induces amplitude-dependent phase shifts (shearing). The emergent blockade effect acts as a rigorous constraint on the photon number distribution, preventing the st ate from spreading indefinitely. It is precisely this dynamical balance—confinement via blockade against shearing via nonlinearity—that sculpts the target non-classical state, in close analogy with the dynamic discussed in Ref.~\cite{Raimond_2010}.

Owing to the quasi-periodic nature of these dynamics, the system approaches the target state at multiple interaction times, resulting in distinct solution branches. We find that the highest fidelities are achieved in the regime
$|\alpha|<1$ (see Fig. \ref{fig:results2}(c)).

Although the nonlinear nature of the system prevents a complete analytical solution, we find analytic results for the PACS generations in the weak drive limit using a perturbative method (see Supplemental Material \cite{SM}).

Finally, although we employed the remnant photon number $n_r$, specifically for PACS recognition, its utility extends to the characterization of general non-Gaussian resources. Because $n_r$ quantifies non-classicality independent of the quadrature definition, it is superior to fixed-phase metrics for analyzing dynamic systems.

{\it Conclusions.---} We present a post-selection free protocol for generating photon-added coherent states (PACS) with high multi-photon addition. The protocol is based on a driven Kerr nonlinear medium inside a cavity, which is a well-known platform. The generation of multi-photon PACS is heralded by first identifying the formation of coherently displaced PACS, namely, $ \ket {\Gamma_{\alpha,m}}$ in Eq. (\ref{eq:PACS}).  We proposed a displacement-independent witness that recognizes when a high-fidelity PACS can emerge during evolution, $n_r$ in Eq. \eqref{eq:meas}.

The results show the deterministic generation of PACS with at least 10 added photons (See Supplementary Material \cite{SM}) and fidelities exceeding $99\%$ under unitary evolution. Crucially, the rapid timescale of our generation protocol naturally mitigates the impact of environmental decoherence, preserving the non-classical features of the field. We further show that high-fidelity multi-photon PACS remain accessible even in the presence of realistic dissipation, using parameters within current state-of-the-art technology. By overcoming the probabilistic bottlenecks of previous approaches, our results unlock a deterministic pathway toward complex non-Gaussian states for quantum metrology and continuous-variable information processing.

{\it Acknowledgments.---} This work was supported in part by FONDECYT Grants No. 1230897, No. 1240204. ANID Doctoral Fellowship Grant No. 21232285, and ANID - Millennium Science Initiative Program Grant No. ICN17-012. R.G-J. gratefully acknowledges support from PAPIIT-UNAM (Grants No. IA103024 and IA105926) and SECIHTI (Award No. CBF-2025-I-1090).

\bibliography{main}

@article{Agarwal_1991,
  title = {Nonclassical properties of states generated by the excitations on a coherent state},
  author = {Agarwal, G. S. and Tara, K.},
  journal = {Phys. Rev. A},
  volume = {43},
  issue = {1},
  pages = {492--497},
  numpages = {0},
  year = {1991},
  month = {Jan},
  publisher = {American Physical Society},
  doi = {10.1103/PhysRevA.43.492},
  url = {https://link.aps.org/doi/10.1103/PhysRevA.43.492}
}

@article{Drummond_1980,
doi = {10.1088/0305-4470/13/2/034},
url = {https://dx.doi.org/10.1088/0305-4470/13/2/034},
year = {1980},
month = {feb},
publisher = {},
volume = {13},
number = {2},
pages = {725},
author = {P D Drummond and  D F Walls},
title = {Quantum theory of optical bistability. I. Nonlinear polarisability model},
journal = {Journal of Physics A: Mathematical and General}
}

@article{Imamo_1997,
  title = {Strongly Interacting Photons in a Nonlinear Cavity},
  author = {Imamo\ifmmode \bar{g}\else \={g}\fi{}lu, A. and Schmidt, H. and Woods, G. and Deutsch, M.},
  journal = {Phys. Rev. Lett.},
  volume = {79},
  issue = {8},
  pages = {1467--1470},
  numpages = {0},
  year = {1997},
  month = {Aug},
  publisher = {American Physical Society},
  doi = {10.1103/PhysRevLett.79.1467},
  url = {https://link.aps.org/doi/10.1103/PhysRevLett.79.1467}
}

@article{Braunstein2005,
  title = {Quantum information with continuous variables},
  author = {Braunstein, Samuel L. and van Loock, Peter},
  journal = {Rev. Mod. Phys.},
  volume = {77},
  issue = {2},
  pages = {513--577},
  numpages = {0},
  year = {2005},
  month = {Jun},
  publisher = {American Physical Society},
  doi = {10.1103/RevModPhys.77.513},
  url = {https://link.aps.org/doi/10.1103/RevModPhys.77.513}
}

@article{Sperling2014,
  title = {Quantum state engineering by click counting},
  author = {Sperling, J. and Vogel, W. and Agarwal, G. S.},
  journal = {Phys. Rev. A},
  volume = {89},
  issue = {4},
  pages = {043829},
  numpages = {10},
  year = {2014},
  month = {Apr},
  publisher = {American Physical Society},
  doi = {10.1103/PhysRevA.89.043829},
  url = {https://link.aps.org/doi/10.1103/PhysRevA.89.043829}
}

@article{Zavatta2004,
author = {Alessandro Zavatta  and Silvia Viciani  and Marco Bellini },
title = {Quantum-to-Classical Transition with Single-Photon-Added Coherent States of Light},
journal = {Science},
volume = {306},
number = {5696},
pages = {660-662},
year = {2004},
doi = {10.1126/science.1103190},
URL = {https://www.science.org/doi/abs/10.1126/science.1103190}}

@article{Pizzimenti2021,
  title = {Non-Gaussian photonic state engineering with the quantum frequency processor},
  author = {Pizzimenti, Andrew J. and Lukens, Joseph M. and Lu, Hsuan-Hao and Peters, Nicholas A. and Guha, Saikat and Gagatsos, Christos N.},
  journal = {Phys. Rev. A},
  volume = {104},
  issue = {6},
  pages = {062437},
  numpages = {13},
  year = {2021},
  month = {Dec},
  publisher = {American Physical Society},
  doi = {10.1103/PhysRevA.104.062437},
  url = {https://link.aps.org/doi/10.1103/PhysRevA.104.062437}
}

@article{Nielsen2007,
  title = {Photon number states generated from a continuous-wave light source},
  author = {Nielsen, Anne E. B. and M\o{}lmer, Klaus},
  journal = {Phys. Rev. A},
  volume = {75},
  issue = {4},
  pages = {043801},
  numpages = {8},
  year = {2007},
  month = {Apr},
  publisher = {American Physical Society},
  doi = {10.1103/PhysRevA.75.043801},
  url = {https://link.aps.org/doi/10.1103/PhysRevA.75.043801}
}

@article{Miranowicz2013,
  title = {Two-photon and three-photon blockades in driven nonlinear systems},
  author = {Miranowicz, Adam and Paprzycka, Ma\l{}gorzata and Liu, Yu-xi and Bajer, Ji\ifmmode \check{r}\else \v{r}\fi{}\'{\i} and Nori, Franco},
  journal = {Phys. Rev. A},
  volume = {87},
  issue = {2},
  pages = {023809},
  numpages = {10},
  year = {2013},
  month = {Feb},
  publisher = {American Physical Society},
  doi = {10.1103/PhysRevA.87.023809},
  url = {https://link.aps.org/doi/10.1103/PhysRevA.87.023809}
}

@article{Anikin_2021,
  title = {Multiphoton resonance in a driven Kerr oscillator in the presence of high-order nonlinearities},
  author = {Anikin, Evgeny V. and Maslova, Natalya S. and Gippius, Nikolay A. and Sokolov, Igor M.},
  journal = {Phys. Rev. A},
  volume = {104},
  issue = {5},
  pages = {053106},
  numpages = {9},
  year = {2021},
  month = {Nov},
  publisher = {American Physical Society},
  doi = {10.1103/PhysRevA.104.053106},
  url = {https://link.aps.org/doi/10.1103/PhysRevA.104.053106}
}

@article{Fadrny_2024, 
title={Experimental preparation of multiphoton-added coherent states of light}, 
volume={10}, 
DOI={10.1038/s41534-024-00885-y}, 
number={1}, journal={npj Quantum Information}, 
author={Fadrný, Jiří and Neset, Michal and Bielak, Martin and Ježek, Miroslav and Bílek, Jan and Fiurášek, Jaromír}, year={2024}, month=sep }

@article{Kumar2013,
  title = {Experimental Characterization of Bosonic Creation and Annihilation Operators},
  author = {Kumar, R. and Barrios, E. and Kupchak, C. and Lvovsky, A. I.},
  journal = {Phys. Rev. Lett.},
  volume = {110},
  issue = {13},
  pages = {130403},
  numpages = {5},
  year = {2013},
  month = {Mar},
  publisher = {American Physical Society},
  doi = {10.1103/PhysRevLett.110.130403},
  url = {https://link.aps.org/doi/10.1103/PhysRevLett.110.130403}
}

@article{Barbieri2010,
  title = {Non-Gaussianity of quantum states: An experimental test on single-photon-added coherent states},
  author = {Barbieri, Marco and Spagnolo, Nicol\`o and Genoni, Marco G. and Ferreyrol, Franck and Blandino, R\'emi and Paris, Matteo G. A. and Grangier, Philippe and Tualle-Brouri, Rosa},
  journal = {Phys. Rev. A},
  volume = {82},
  issue = {6},
  pages = {063833},
  numpages = {5},
  year = {2010},
  month = {Dec},
  publisher = {American Physical Society},
  doi = {10.1103/PhysRevA.82.063833},
  url = {https://link.aps.org/doi/10.1103/PhysRevA.82.063833}
}

@article{Uria2020,
  title = {Deterministic Generation of Large Fock States},
  author = {Uria, M. and Solano, P. and Hermann-Avigliano, C.},
  journal = {Phys. Rev. Lett.},
  volume = {125},
  issue = {9},
  pages = {093603},
  numpages = {6},
  year = {2020},
  month = {Aug},
  publisher = {American Physical Society},
  doi = {10.1103/PhysRevLett.125.093603},
  url = {https://link.aps.org/doi/10.1103/PhysRevLett.125.093603}
}

@article{Iyama_2024, 
title={Observation and manipulation of quantum interference in a superconducting Kerr parametric oscillator}, 
volume={15}, 
DOI={10.1038/s41467-023-44496-1}, 
number={1}, 
journal={Nature Communications}, 
author={Iyama, Daisuke and Kamiya, Takahiko and Fujii, Shiori and Mukai, Hiroto and Zhou, Yu and Nagase, Toshiaki and Tomonaga, Akiyoshi and Wang, Rui and Xue, Jiao-Jiao and Watabe, Shohei and Kwon, Sangil and Tsai, Jaw-Shen}, year={2024}, month=jan }

@article{Sivakumar2014, 
title={Truncated Coherent States and Photon-Addition}, 
volume={53}, 
DOI={10.1007/s10773-013-1967-7}, 
number={5}, 
journal={International Journal of Theoretical Physics}, 
author={Sivakumar, S.}, year={2014}, 
url ={ https://doi.org/10.1007/s10773-013-1967-7}}

@article{Hoffman2011,
  title = {Dispersive Photon Blockade in a Superconducting Circuit},
  author = {Hoffman, A. J. and Srinivasan, S. J. and Schmidt, S. and Spietz, L. and Aumentado, J. and T\"ureci, H. E. and Houck, A. A.},
  journal = {Phys. Rev. Lett.},
  volume = {107},
  issue = {5},
  pages = {053602},
  numpages = {4},
  year = {2011},
  month = {Jul},
  publisher = {American Physical Society},
  doi = {10.1103/PhysRevLett.107.053602},
  url = {https://link.aps.org/doi/10.1103/PhysRevLett.107.053602}
}

@article{Glauber1963, 
title={The Quantum Theory of Optical Coherence}, 
volume={130}, 
DOI={10.1103/physrev.130.2529}, 
number={6}, 
journal={Physical Review}, 
author={Glauber, Roy J.}, 
year={1963},
month=jun, pages={2529–2539} 
}

@book{Agarwal2012, title={Quantum Optics}, 
url={https://doi.org/10.1017/cbo9781139035170}, 
publisher={Cambridge University Press}, 
author={Agarwal, Girish S.}, 
year={2012}, month=nov }

@article{Hillery_1989, 
title={Total noise and nonclassical states}, 
volume={39}, 
DOI={10.1103/physreva.39.2994}, 
number={6}, 
journal={Physical Review A}, 
author={Hillery, Mark}, 
year={1989}, month=mar, 
pages={2994–3002} }

@article{Yadin_2018, 
title={Operational Resource Theory of Continuous-Variable Nonclassicality}, volume={8}, 
DOI={10.1103/physrevx.8.041038}, 
number={4}, 
journal={Physical Review X}, 
author={Yadin, Benjamin and Binder, Felix C. and Thompson, Jayne and Narasimhachar, Varun and Gu, Mile and Kim, M. S.}, 
year={2018}, month=dec }

@article{Tan_2019, 
title={Nonclassical light and metrological power: An introductory review}, 
volume={1}, DOI={10.1116/1.5126696}, 
number={1}, 
journal={AVS Quantum Science}, 
author={Tan, Kok Chuan and Jeong, Hyunseok}, 
year={2019}, month=nov }

@book{Haroche_2006, 
title={Exploring the Quantum: Atoms, Cavities, and Photons},
ISBN={9780191523243}, 
publisher={OUP Oxford}, 
author={Haroche, Serge and Raimond, Jean-Michel}, 
year={2006}, month=aug }

@article{Cahill_1969, 
title={Ordered Expansions in Boson Amplitude Operators}, 
volume={177}, 
DOI={10.1103/physrev.177.1857}, 
number={5}, 
journal={Physical Review}, 
author={Cahill, K. E. and Glauber, R. J.}, 
year={1969}, month=jan, pages={1857–1881} }

@article{Laiho_2009, 
title={Direct probing of the Wigner function by time-multiplexed detection of photon statistics}, 
volume={11}, 
DOI={10.1088/1367-2630/11/4/043012}, 
number={4}, journal={New Journal of Physics}, 
author={Laiho, K and Avenhaus, M and Cassemiro, K N and Silberhorn, Ch}, 
year={2009}, month=apr, pages={043012} }

@article{Solano_2005, 
title={Selective interactions in trapped ions: State reconstruction and quantum logic}, 
volume={71}, DOI={10.1103/physreva.71.013813}, 
number={1}, journal={Physical Review A}, 
author={Solano, E.}, 
year={2005}, month=jan }

@article{Raimond_2010, 
title={Phase Space Tweezers for Tailoring Cavity Fields by Quantum Zeno Dynamics}, 
volume={105}, DOI={10.1103/physrevlett.105.213601}, 
number={21}, journal={Physical Review Letters}, 
author={Raimond, J. M. and Sayrin, C. and Gleyzes, S. and Dotsenko, I. and Brune, M. and Haroche, S. and Facchi, P. and Pascazio, S.}, 
year={2010}, month=nov }

@article{Pinheiro_2012, 
title={Quantum communication with photon-added coherent states}, 
volume={12}, DOI={10.1007/s11128-012-0400-0}, 
number={1}, journal={Quantum Information Processing}, 
author={Pinheiro, Paulo Vinícius Pereira and Ramos, Rubens Viana}, 
year={2012}, month=mar, pages={537–547} }

@article{Chatterjee_2021, 
title={Quantifying Quantum Correlation of Quasi‐Werner State and Probing Its Suitability for Quantum Teleportation},
volume={533}, DOI={10.1002/andp.202100201}, 
number={10}, journal={Annalen der Physik}, 
author={Chatterjee, Arpita and Thapliyal, Kishore and Pathak, Anirban}, 
year={2021}, month=jul }

@article{Hahn_2022, 
title={Deterministic Gaussian conversion protocols for non-Gaussian single-mode resources}, 
volume={105}, DOI={10.1103/physreva.105.062446}, 
number={6}, journal={Physical Review A}, 
author={Hahn, Oliver and Holmvall, Patric and Stadler, Pascal and Ferrini, Giulia and Ferraro, Alessandro}, 
year={2022}, month=jun }

@misc{SM,
  note = "See Supplemental Material at
    URL-will-be-inserted-by-publisher for more details."
}

@article{Kudra_2025, 
title={Experimental realization of deterministic and selective photon addition in a bosonic mode assisted by an ancillary qubit}, 
volume={10}, DOI={10.1088/2058-9565/ae0519}, 
number={4}, journal={Quantum Science and Technology}, 
author={Kudra, Marina and Jirlow, Martin and Kervinen, Mikael and Eriksson, Axel M and Quijandría, Fernando and Delsing, Per and Abad, Tahereh and Gasparinetti, Simone}, 
year={2025}, month=sep, pages={045037} }

@article{Kudra_2022, 
 title={Robust Preparation of Wigner-Negative States with Optimized SNAP-Displacement Sequences}, 
 volume={3}, DOI={10.1103/prxquantum.3.030301}, 
 number={3}, journal={PRX Quantum}, 
 author={Kudra, Marina and Kervinen, Mikael and Strandberg, Ingrid and Ahmed, Shahnawaz and Scigliuzzo, Marco and Osman, Amr and Lozano, Daniel Pérez and Tholén, Mats O. and Borgani, Riccardo and Haviland, David B. and Ferrini, Giulia and Bylander, Jonas and Kockum, Anton Frisk and Quijandría, Fernando and Delsing, Per and Gasparinetti, Simone}, 
 year={2022}, month=jul }

@article{Heeres_2015, 
title={Cavity State Manipulation Using Photon-Number Selective Phase Gates}, 
volume={115}, DOI={10.1103/physrevlett.115.137002}, 
number={13}, journal={Physical Review Letters}, 
author={Heeres, Reinier W. and Vlastakis, Brian and Holland, Eric and Krastanov, Stefan and Albert, Victor V. and Frunzio, Luigi and Jiang, Liang and Schoelkopf, Robert J.}, 
year={2015}, month=sep }

\clearpage
\onecolumngrid

\setcounter{page}{1}

\setcounter{section}{0}
\setcounter{equation}{0}
\setcounter{figure}{0}
\renewcommand{\theequation}{S\arabic{equation}}
\renewcommand{\thefigure}{S\arabic{figure}}

\begin{center}
    \textbf{\Large SUPPLEMENTARY MATERIAL: Post-Selection Free Generation of Multi-Photon Added Coherent States}
\end{center}

\section{Remnant photon number}

In this section, we provide the derivation of the {\it remnant photon number}. Consider a general bosonic state $\ket{\psi}$ and a complex displacement parameter $z = x + i y$. The mean photon number after applying the displacement $\hat{D}(z)$ is
\begin{equation}
    \begin{split}
n(z)&=\bra{\psi}\hat{D}^\dagger(z)\hat{n}\hat{D}(z)\ket{\psi}\\
&=x^2+y^2+2x\left<\hat{x}\right>+2y\left<\hat{p}\right>+\left<\hat{x}^2\right>+\left<\hat{p}^2\right>-1/2,
    \end{split}
\end{equation}
where $\hat{x}=(\hat{a}+\hat{a}^\dagger)/2$ and $\hat{p}=(\hat{a}-\hat{a}^\dagger)/2i$ are the field quadrature operators. Minimizing $n(z)$ with respect to $x$ and $y$ yields the critical point $x_c=-\left<\hat{x}\right>, y_c=-\left<\hat{p}\right>$, corresponding to a displacement that shifts the state to the origin of phase space. Substituting these values back into $n(z)$, we obtain the remnant photon number
\begin{equation}
    n_r=\left<\hat{x}^2\right>-\left<\hat{x}\right>^2+\left<\hat{p}^2\right>-\left<\hat{p}\right>^2-1/2,
\end{equation}
which quantifies the residual photon number associated with intrinsic quantum fluctuations after optimal displacement.
\\

In the case of a general quadrature $\hat{x}_\theta=\hat{x}\cos\theta+\hat{p}\sin\theta$, one can verify the conserved quantities
\begin{equation}
\langle \hat{x}_\theta^2 \rangle + \langle \hat{x}_{\theta+\pi/2}^2 \rangle
= \langle \hat{x}^2 \rangle + \langle \hat{p}^2 \rangle,
\end{equation}
reflecting energy conservation under phase-space rotations, and
\begin{equation}
\langle \hat{x}_\theta \rangle^2 + \langle \hat{x}_{\theta+\pi/2} \rangle^2
= \langle \hat{x} \rangle^2 + \langle \hat{p} \rangle^2,
\end{equation}
corresponding to the invariance of the mean amplitude under phase-space rotations. Combining these results, the remnant photon number takes the form
\begin{equation}
n_r=\left<\hat{x}_\theta^2\right>-\left<\hat{x}_\theta\right>^2+\langle\hat{x}_{\theta+\pi/2}^2\rangle-\langle\hat{x}_{\theta+\pi/2}\rangle^2-1/2,
\end{equation}
which shows explicitly that $n_r$ is invariant under the phase-space direction in which it is probed, as discussed in the main text. This invariance highlights that $n_r$ is a basis-independent quantity, determined solely by intrinsic quantum fluctuations. Consequently, it provides a physically meaningful ordering of non-classicality.

\section{Photon-number matching condition}

As discussed in the main text, the state $\ket{\Psi(t)}=\hat D(\zeta)\hat D(-\gamma)\ket{\psi(t)}$ resembles the photon added coherent state (PACS) $\ket{\alpha,n}$, where $\gamma=\langle \hat x\rangle+i\langle \hat p\rangle$. We therefore impose, as a physically motivated constraint, that the displaced state $\hat D(-\gamma)\ket{\psi(t)}$ at the relevant evolution time and the displaced PACS $\hat D(-\zeta)\ket{\alpha,n}$ should have the same mean number of photons. Writing $\alpha=|\alpha|e^{i\phi}$ and $\zeta=|\zeta|e^{i\phi}$ so that both amplitudes share the same phase $\phi$, this condition reads
\begin{equation}
\begin{aligned}
\bra{\psi(t)}\hat D(\gamma)\hat n\hat D^\dagger(\gamma)\ket{\psi(t)}&=n_r\bigl(\ket{\psi(t)}\bigr)\\
&=\bra{\alpha,n}\hat D(\zeta)\hat n\hat D^\dagger(\zeta)\ket{\alpha,n}\\
&=\bra{\alpha,n}\hat{a}\hat{a}^\dagger-1 -|\zeta|(\hat{a} e^{-i\phi}+\hat{a}^{\dagger} e^{i\phi})+|\zeta|^2\ket{\alpha,n}\\
&=\frac{N_{n+1}}{N_n}-2|\zeta|\langle \hat{x}_\phi\rangle_{\alpha,n}+|\zeta|^2-1
\label{eq:n}
\end{aligned}
\end{equation}
where $\langle \hat{x}_\phi\rangle_{\alpha,n}=|\alpha|L_n^{(1)}(-|\alpha|^2)/L_n(-|\alpha|^2)$ with $L_n^{(1)}(x)$ and $L_n(x)$ denoting the associated and ordinary Laguerre polynomials \cite{Cahill_1969}, respectively. This expression follows from the results derived in Ref.~\cite{Agarwal_1991} under the condition $\arg(\alpha)=\arg(\zeta)=\phi$. 

\begin{figure}[H]
    \centering
    \includegraphics[width=0.7\linewidth]{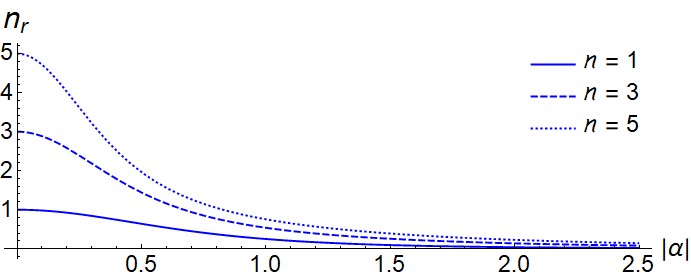}
    \caption{Remnant photon number $n_r$ of the photon-added coherent state $\ket{\alpha,n}$ as a function of the coherent amplitude $|\alpha|$ for different photon-addition numbers $n$. At $|\alpha|=0$, the remnant photon number coincides with the photon number of the corresponding Fock state. As $|\alpha|$ increases, $n_r$ decreases monotonically and approaches zero in the large amplitude limit, where the state resembles a coherent state. Larger values of $n$ yield higher remnant photon numbers, reflecting the enhanced nonclassicality induced by photon addition.}
    \label{fig:nr}
\end{figure}

Equation~\eqref{eq:n} can also be used to determine the remnant photon number of the PACS itself. Minimizing the right-hand side with respect to $|\zeta|$ yields the optimal displacement $
|\zeta|_{opt}=\langle \hat x_\phi\rangle_{\alpha,n}$, from which we obtain
\begin{equation}
n_r(\ket{\alpha,n})=\frac{N_{n+1}}{N_n}-\langle \hat x_\phi\rangle_{\alpha,n}^2-1,
\end{equation}
whose behavior can be seen in Fig.~\ref{fig:nr}. Finally, solving Eq.~\eqref{eq:n} for $|\zeta|$ gives
\begin{equation}
\begin{aligned}
|\zeta|
&=
\langle \hat x_\phi\rangle_{\alpha,n}
\pm
\sqrt{
\langle \hat x_\phi\rangle_{\alpha,n}^2
-
\frac{N_{n+1}}{N_n}
+
1
+
n_r\bigl(\ket{\psi(t)}\bigr)
}
\\
&=
\langle \hat x_\phi\rangle_{\alpha,n}
\pm
\sqrt{
n_r\bigl(\ket{\psi(t)}\bigr)
-
n_r(\ket{\alpha,n})
}.
\end{aligned}
\end{equation}
This expression establishes the relation between the displacement amplitude $|\zeta|$ and the difference between the remnant photon numbers of the evolved state and the corresponding PACS.

\section{Physical Interpretation of the Condition $\delta=2\chi$}

The condition $\delta = 2\chi$ plays a central role in the emergence of PACS-like structures as a spectral constraint that preserves the intrinsic phase organization of the Kerr system. This condition is satisfied when the field linear evolution frequency $\delta$ coincides with the smallest Kerr evolution frequency $2\chi$. As a consequence, the periodicity of the linear evolution matches the periodicity of the nonlinear evolution.

The effective Fock-state energies of the driven Kerr Hamiltonian are
\begin{equation}
E_n = \hbar \delta n + \hbar \chi n(n-1),
\end{equation}
with level spacings
\begin{equation}
E_{n+1}-E_n = \hbar(\delta + 2\chi n).
\end{equation}
For zero detuning ($\delta=0$), the states $\ket{0}$ and $\ket{1}$ are degenerate. The condition $\delta=2\chi$ removes this low-energy degeneracy while preserving the characteristic quadratic Kerr structure, yielding
\begin{equation}
E_n=\hbar\chi n(n+1).
\end{equation}
The dynamics retains the intrinsic Kerr structure, but with a reordered low-energy ladder free from the $\ket{0}$--$\ket{1}$ degeneracy.

The coherent drive,
\begin{equation}
\hat H_{drive}=\hbar\mathcal{E}(\hat a+\hat a^\dagger),
\end{equation}
induces sequential transitions between neighboring Fock states with a given phase. The resulting dynamics is therefore governed by the interplay between driven transitions and Kerr-induced phase evolution. Since the condition $\delta=2\chi$ preserves the monotonic Kerr ladder structure, the driven transitions can interfere constructively over time, favoring partial rephasing of low-lying Fock components and the emergence of structured nonclassical states.

\begin{figure}
    \centering
    \includegraphics[width=0.6\linewidth]{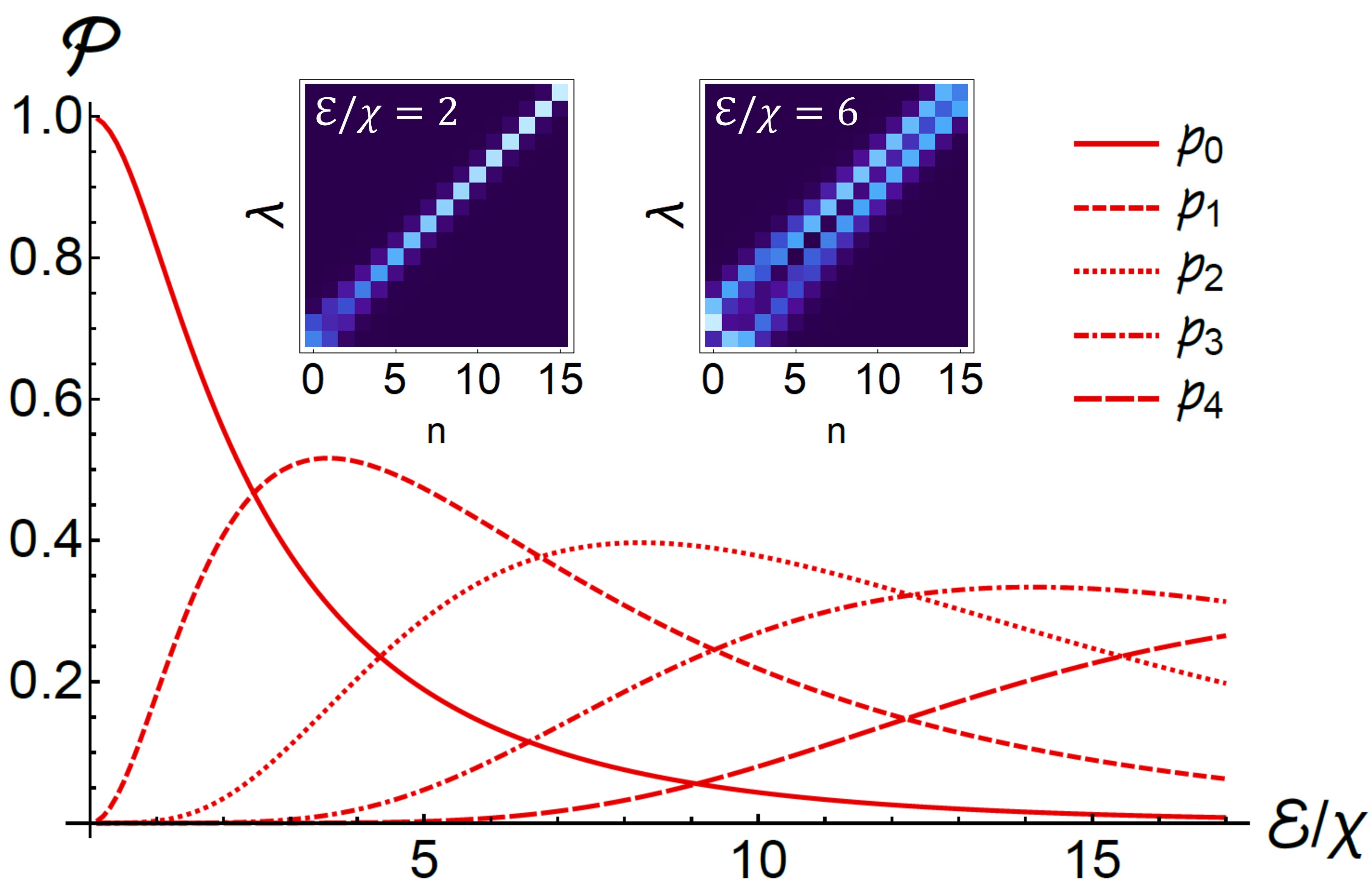}
    \caption{Spectral decomposition probabilities $p_n=|\langle \lambda_n|0\rangle|^2$ of the vacuum state projected onto the eigenstates $\ket{\lambda_n}$ of the Hamiltonian $\mathcal{H}_R$ as a function of the ratio $\mathcal{E}/\chi$. In the weak-driving regime, the vacuum state is predominantly composed of the lowest-energy eigenstates. As the ratio $\mathcal{E}/\chi$ increases, a larger number of eigenstates contributes significantly to the decomposition, resulting in a broader spectral distribution. The insets show the Fock-basis representation of the eigenstates for $\mathcal{E}/\chi=2$ and $\mathcal{E}/\chi=6$, respectively. For weak driving, the eigenstates remain strongly localized near the diagonal, whereas stronger driving leads to a broader distribution over Fock states, reflecting the increasing complexity of the dressed-state structure.
}
    \label{fig:prob}
\end{figure}

\section{Spectral and Dynamical Properties of the System}
In the main text, the Hamiltonian governing the time evolution of the system is given by
\begin{equation}
\mathcal{H}=\hbar \omega_0 \hat{a}^{\dagger}\hat{a} +\hbar \mathcal{E}\left( \hat{a}\ e^{i\omega_d t} + \hat{a}^{\dagger} e^{-i\omega_d t} \right) + \hbar\chi \hat{a}^{\dagger 2}\hat{a}^2,
\end{equation}
where $\omega_0$ is the cavity frequency, $\omega_d$ is the driving frequency, $\mathcal{E}$ denotes the driving amplitude and $\chi$ characterizes the Kerr nonlinearity. Transforming into the rotating frame defined by $R(t)=e^{-i\omega_d \hat{a}^\dagger\hat{a} t}$, the Hamiltonian becomes 
\begin{equation}
\mathcal{H}_R
= R \,\mathcal{H} R^{-1}+i\hbar (\partial_t R)R^{-1}=
\hbar\delta\,\hat{a}^{\dagger}\hat{a}
+
\hbar\mathcal{E}
\left(
\hat{a}
+
\hat{a}^{\dagger}
\right)
+
\hbar\chi\,\hat{a}^{\dagger 2}\hat{a}^2,
\label{eq:H}
\end{equation}
where $\delta=\omega_0-\omega_d=2\chi$ is the selected detuning between the cavity and driving frequencies. Due to the nonlinear nature of the system, analytical expressions for the eigenstates are generally not available. Denoting the eigenstates and eigenvalues of $\mathcal{H}_R$ by $\ket{\lambda_j}$ and $\lambda_j$, respectively, the initial vacuum state can be expanded as
\begin{equation}
\ket{\psi(0)}=\ket{0}=\sum_j c_j \ket{\lambda_j}.
\end{equation}
As shown in Fig.~\ref{fig:prob}, the vacuum state is represented by a finite superposition of these eigenstates with probabilities $p_j=|c_j|^2$. As the ratio $\mathcal{E}/\chi$ increases, a larger number of eigenstates contributes significantly to the decomposition. In the weak-driving regime, the dominant eigenstates themselves involve only a few Fock components, typically no more than three. In contrast, for larger values of $\mathcal{E}/\chi$, the structure of the eigenstates becomes increasingly delocalized in the Fock basis, as illustrated in the insets of Fig.~\ref{fig:prob}.

Moreover, since the Hamiltonian $\mathcal{H}_R$ is represented by a real symmetric matrix in the Fock basis, its eigenvalues $\lambda_j$ are real and its eigenvectors $\ket{\lambda_j}$ can be chosen with purely real coefficients. Consequently, the overlaps $c_j=\braket{\lambda_j}{0}$ are also real quantities. The state of the system at time $t$ is therefore given by
\begin{equation}
    \ket{\psi(t)}=\sum_j \braket{\lambda_j}{0} e^{-i\lambda_j t} \ket{\lambda_j}.
\end{equation}
The expectation value of the photon number operator then reads
\begin{equation}
\begin{aligned}
\langle \hat n(t)\rangle
&=
\sum_{j,k}
\braket{0}{\lambda_k}
\braket{\lambda_j}{0}
e^{i(\lambda_k-\lambda_j)t/\hbar}
\bra{\lambda_k}\hat n\ket{\lambda_j}
\\
&=
\sum_j
\abs{\braket{0}{\lambda_j}}^2
\bra{\lambda_j}\hat n\ket{\lambda_j}
+ \sum_{j\neq k}
\braket{0}{\lambda_k}
\braket{\lambda_j}{0}
\bra{\lambda_k}\hat n\ket{\lambda_j}
\cos\!\left[
(\lambda_k-\lambda_j)t
\right].
\end{aligned}
\end{equation}
Here we have used the fact that the matrix elements $\bra{\lambda_k}\hat n\ket{\lambda_j}$ are real. The first term describes the static contribution arising from the populations of the dressed eigenstates, while the second term encodes coherent oscillations generated by quantum interference between different eigenstates. The oscillation frequencies are determined by the energy differences $(\lambda_k-\lambda_j)$, revealing the spectral structure of the driven nonlinear system.

\begin{figure}
    \centering
    \includegraphics[width=\linewidth]{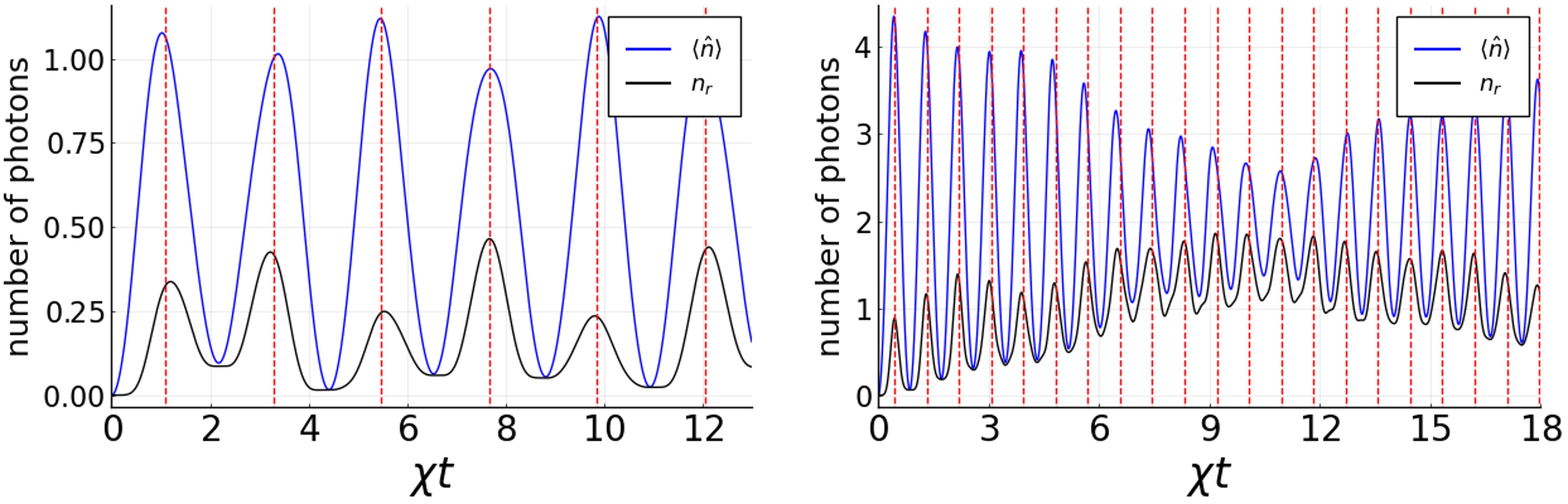}
    \caption{Time evolution of the mean photon number $\langle \hat n(t)\rangle$ (blue curves) and the remnant photon number $n_r$ (black curves) for two different driving regimes $\mathcal{E}/\chi=\{1.5,6.5\}$. The vertical dashed red lines indicate the characteristic oscillation period determined by the dominant energy splitting $(\lambda_m-\lambda_n)$ arising from the interference between the two most relevant dressed eigenstates in the spectral decomposition of the initial vacuum state. In the weak-driving regime (left panel), the dynamics are governed by a small number of eigenstates, leading to nearly periodic oscillations with low photon occupation and a comparatively small remnant photon number. In the stronger-driving regime (right panel), a larger number of dressed eigenstates contributes to the dynamics, producing higher photon populations and more complex oscillatory behavior. The remnant photon number remains systematically smaller than $\langle \hat n(t)\rangle$, reflecting the removal of the coherent displacement contribution.}
    \label{fig:photon_number}
\end{figure}

A particularly relevant situation occurs when the dynamics are dominated by two eigenstates, $\ket{\lambda_m}$ and $\ket{\lambda_n}$. In this case, the photon-number mean value can be approximated as
\begin{equation}
    \langle \hat n(t)\rangle=p_m
\bra{\lambda_m}\hat n\ket{\lambda_m}+p_n
\bra{\lambda_n}\hat n\ket{\lambda_n}+ 2\braket{0}{\lambda_m}
\braket{\lambda_n}{0}
\bra{\lambda_m}\hat n\ket{\lambda_n}
\cos\!\left[
(\lambda_m-\lambda_n)t
\right].
\label{eq:num}
\end{equation}
The dynamics are therefore characterized by a dominant oscillation period $2\pi/(\lambda_m-\lambda_n)$. This characteristic timescale is highlighted by the vertical red lines in Fig.~\ref{fig:photon_number}. Additionally, it is worth to mention that the system is initially prepared in the vacuum state, then one has $\langle \hat n(0)\rangle=0$. Consequently, the interference contribution must compensate the positive diagonal terms at $t=0$, implying that $\braket{0}{\lambda_m}\braket{\lambda_n}{0}\bra{\lambda_m}\hat n\ket{\lambda_n}<0$. As a result, the photon number reaches its maximum value when $(\lambda_m-\lambda_n)t=\pi$.

\section{Weak-Driving Regime Analysis}

\begin{figure}
    \centering
    \includegraphics[width=\linewidth]{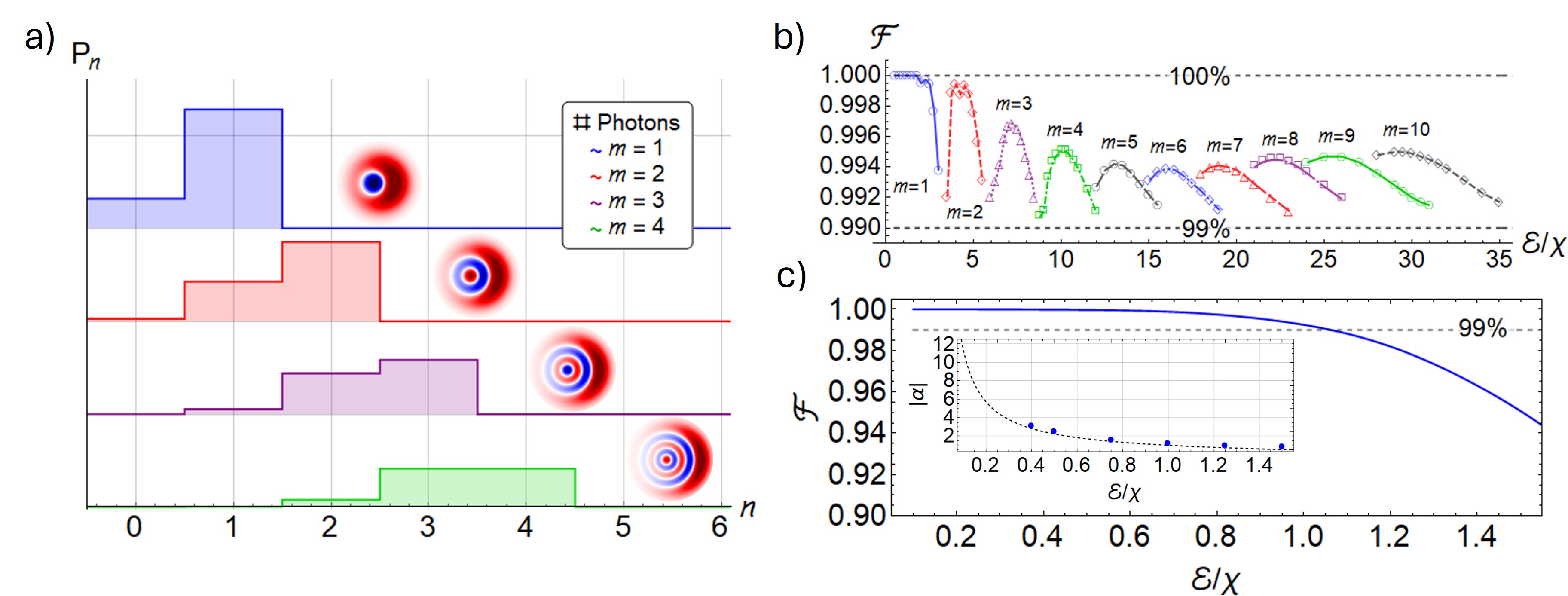}
    \caption{(a) Photon-number distribution $P_n$ of the states $\ket{\Gamma_{\alpha,m}}$ for different photon-addition numbers $m$ and coherent amplitude $\alpha=0.5$. Each distribution is shifted vertically for clarity. As the number of added photons increases, the probability distribution moves toward higher photon-number sectors and becomes strongly concentrated around a characteristic maximum photon number, reflecting the photon-blockade structure induced by the nonlinear excitation process. The insets show the corresponding phase-space structure of the states, highlighting the increasing nonclassical character. (b) Fidelity $\mathcal{F}$ obtained from the numerical optimization between the dynamically generated states and the target photon-added coherent states as a function of the normalized driving strength $\mathcal{E}/\chi$ for photon-addition numbers up to $m=10$ under the resonance condition $\delta=2\chi$. Each curve identifies the parameter region where the dynamics most efficiently reproduces a given photon-added coherent state. The dashed horizontal line marks the $99\%$ fidelity threshold.
(c) Comparison between numerical and analytical results in the weak-driving regime for the single-photon-added coherent state. The blue curve corresponds to a fitted interpolation of the fidelities obtained between the numerically generated states and the analytical prediction as a function of $\mathcal{E}/\chi$. The inset shows the comparison between the numerically extracted coherent amplitude $\alpha$ (blue points) and the corresponding analytical weak-driving prediction (black dashed line).}
    \label{fig:PACS_dist}
\end{figure}

To gain analytical insight into the generation of PACS, we now focus on the weak-driving regime. In this limit, the dynamics is dominated by a small number of low-excitation Fock states, enabling a perturbative description of both the dressed eigenstates and the resulting time evolution. This regime is particularly relevant because it is precisely where the highest fidelities are obtained, as shown in Fig.~\ref{fig:PACS_dist}(b), making it especially suitable for analytical treatment. 

In order to apply perturbation theory, the driving amplitude must satisfy
\begin{equation}
\mathcal{E}\ll \frac{E_{n+1}-E_n}{\hbar}=\delta+2\chi n,
\end{equation}
so that the dressed eigenstates can be approximated as
\begin{equation}
\ket{n'}=|{n^{(0)}}\rangle+|{n^{(1)}}\rangle+|{n^{(2)}}\rangle+\cdots.
\end{equation}
Up to first order, the corrections are given by
\begin{equation}
    \begin{split}
        |{n^{(0)}}\rangle&=\ket{n},\\
        |{n^{(1)}}\rangle&=\mathcal{E}\sum_{k\ne n}\frac{\bra{k}\hat{a}+\hat{a}^\dagger\ket{n}}{(\delta+\chi(n+k-1))(n-k)} \ket{k}=\mathcal{E}\left(\frac{\sqrt{n}}{\delta+2\chi(n-1)}\ket{n-1}-\frac{\sqrt{n+1}}{\delta+2\chi n}\ket{n+1}\right).
    \end{split}
\end{equation}
The first two dressed states provide the dominant contribution to the dynamics, since higher-order states do not contain a vacuum component at first order. They are therefore approximated by
\begin{equation}
\ket{0'}=\frac{1}{\sqrt{Z_0}}\left(\ket{0}-\frac{\mathcal{E}}{\delta}\ket{1}
\right) \quad \text{and} \quad \ket{1'}=\frac{1}{\sqrt{Z_1}}\left(\ket{1}+\frac{\mathcal{E}}{\delta}\ket{0}-\frac{\mathcal{E}\sqrt{2}}{\delta+2\chi}\ket{2}
\right),
\end{equation}
where $Z_0$ and $Z_1$ are their normalization constants. The state of the system at time $t$ can be expanded in the basis $\{\ket{n'}\}$ as $\ket{\psi(t)}=\sum_n \braket{n'}{0} e^{-i\lambda_n t}\ket{n'}$. Within the weak-driving approximation, the dynamics is mainly governed by the two lowest dressed states, yielding $\ket{\psi(t)}=\braket{0'}{0}e^{-i\lambda_0 t}\ket{0'}+\braket{1'}{0} e^{-i\lambda_1t}\ket{1'}$. Substituting the perturbative expressions for $\ket{0'}$ and $\ket{1'}$, the evolved state in the Fock basis becomes
\begin{equation}
    \ket{\psi(t)}=e^{-i\lambda_0t}\left[\left(\frac{1}{Z_0}+\frac{\mathcal{E}^2}{Z_1 \delta^2}e^{-i(\lambda_1-\lambda_0)t}\right)\ket{0} +\left(\frac{\mathcal{E}}{Z_1 \delta}e^{-i(\lambda_1-\lambda_0)t}-\frac{\mathcal{E}}{Z_0 \delta}\right)\ket{1}-\frac{\mathcal{E}^2\sqrt{2}}{Z_1\delta(\delta+2\chi)}e^{-i(\lambda_1-\lambda_0)t}\ket{2}\right]
\end{equation}
As shown in Fig.~\ref{fig:photon_number}, the remnant photon number is strongly correlated with the total photon number, indicating that the nonclassical behavior becomes maximal approximately when the photon population reaches its maximum value. Within the two dressed state approximation showed in Eq.~\eqref{eq:num}, this occurs when $(\lambda_1-\lambda_0)t=\pi$. At this time, the evolved state reduces to
\begin{equation}
    \ket{\psi_\pi}\propto -\left(\frac{1}{Z_0}-\frac{\mathcal{E}^2}{Z_1 \delta^2}\right)\ket{0}+\frac{\mathcal{E}}{\delta}\left(\frac{1}{Z_0}+\frac{1}{Z_1}\right)\ket{1}-\frac{\mathcal{E}^2\sqrt{2}}{Z_1\delta(\delta+2\chi)}\ket{2}.
    \label{eq:main}
\end{equation}

We now derive an analytical approximation for the generated state in the weak-driving regime and we will compare it with the previous expression. The single photon added coherent state (SPACS), introduced in Ref.~\cite{Agarwal_1991}, is defined as
\begin{equation}
\ket{\alpha,1}
=
\frac{\hat a^\dagger \ket{\alpha}}{\sqrt{N_1}},
\qquad
N_1=1+|\alpha|^2,
\end{equation}
where $\ket{\alpha}=\hat D(\alpha)\ket{0}$ is a coherent state. Applying an additional small displacement $\hat D(\gamma)$ gives
\begin{equation}
\begin{aligned}
\hat D(\gamma)\ket{\alpha,1}
&=
\frac{\hat{D}(\alpha)}{\sqrt{1+\alpha^2}}\hat{D}(\gamma)\overbrace{(\alpha\ket{0}+\ket{1})}^{\ket{\Gamma_{\alpha,1}}},
\end{aligned}
\end{equation}
where $\ket{\Gamma_{\alpha,1}}$ is a state only composed of two Fock states, as shown in Fig.\ref{fig:PACS_dist}.Expanding the displaced state up to first order in $\gamma$ yields
\begin{equation}
\hat{D}(\gamma)\ket{\Gamma_{\alpha,1}}
\approx \frac{1}{\sqrt{1+\alpha^2}}\left[
(\alpha-\gamma)\ket0
+
(1+\alpha\gamma)\ket1
+
\gamma\sqrt2\ket2\right].
\label{eq:PACS_expansion}
\end{equation}
Comparing Eq.~\eqref{eq:main} with the displaced SPACS expansion in Eq.~\eqref{eq:PACS_expansion}, we obtain the system
\begin{equation}
    \begin{split}
        \frac{\alpha-\gamma}{\sqrt{1+\alpha^2}}&=\frac{\mathcal{E}^2}{Z_1 \delta^2}-\frac{1}{Z_0}, \\
        \frac{\gamma}{\sqrt{1+\alpha^2}}&=-\frac{\mathcal{E}^2}{Z_1\delta(\delta+2\chi)}, \\
        \frac{1+\alpha\gamma}{\sqrt{1+\alpha^2}}&=\frac{\mathcal{E}}{\delta}\left(\frac{1}{Z_0}+\frac{1}{Z_1}\right).
    \end{split}
\end{equation}
Assuming $Z_1\simeq Z_0$ and $\delta=2\chi$,  the system can be easily solved, yielding
\begin{equation}
\boxed{
\alpha=-\frac{1-s}{\sqrt{3s(s+2)}}} \quad \text{and}
\quad \boxed{\gamma=\frac{\alpha s}{1-s}},
\end{equation}
where $s=(\mathcal{E}/\chi)^2/8$. These relations establish a direct correspondence between the driving strength and the effective PACS parameters, showing that the weak-driving regime reproduces a displaced SPACS structure. This accounts for the high fidelities in Fig.~\ref{fig:PACS_dist}(b) and is quantitatively confirmed by the agreement with numerics in Fig.~\ref{fig:PACS_dist}(c).

\section{Experimental robustness}

In realistic experimental implementations, imperfections in the control parameters may affect the quality of the generated states. In particular, deviations in the evolution time and in the coherent displacement applied to reconstruct the target PACS can reduce the final fidelity. To quantify the robustness of the protocol, we analyze the fidelity landscape around the optimal preparation parameters.

\begin{figure}[H]
    \centering
    \includegraphics[width=0.95\linewidth]{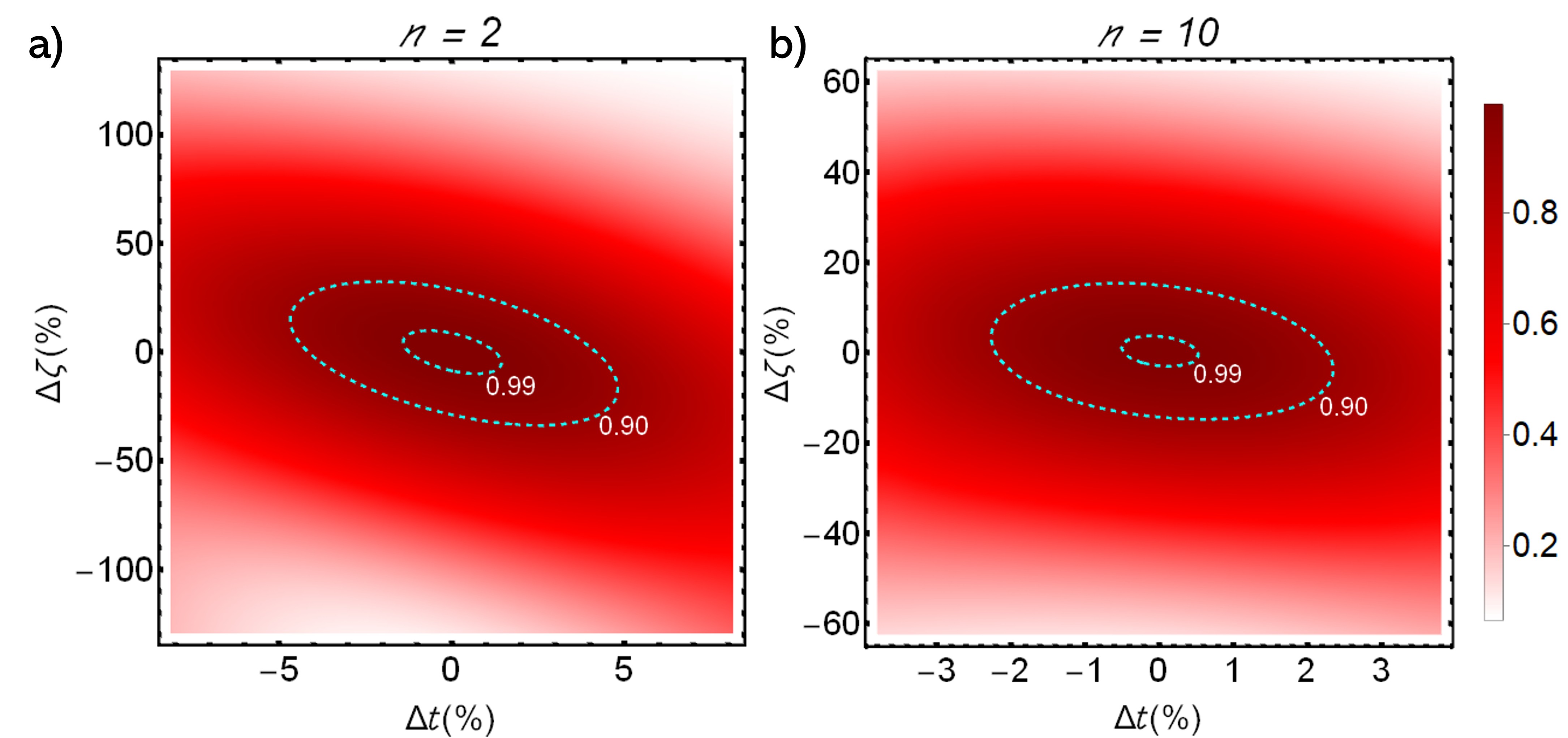}
    \caption{Density plots of the fidelity $\mathcal{F}$ as a function of relative deviations in the displacement parameter $\Delta \zeta$ and evolution time $\Delta t$ for (a) $n=2$ and (b) $n=10$. The reference point $(\Delta t,\Delta \zeta)=(0,0)$ corresponds to the optimal parameters maximizing the fidelity. The dashed cyan contours indicate fidelity thresholds of $0.90$ and $0.99$. In both cases, the fidelity remains high within a finite parameter region, demonstrating the robustness of the generated states against small deviations in the preparation protocol.}
    \label{fig:var}
\end{figure}

Figure~\ref{fig:var} shows the fidelity $\mathcal{F}$ between the generated state and the target photon-added coherent state as a function of relative variations in the displacement parameter, $\Delta\zeta$, and in the evolution time, $\Delta t$. As the photon-addition number increases, the high-fidelity region becomes progressively narrower. This behavior reflects the enhanced sensitivity of higher-order PACS to small perturbations in the preparation protocol. Since these states involve increasingly structured quantum interference between several Fock components, slight deviations in the accumulated dynamical phases or in the displacement amplitude can noticeably modify the resulting state.

\end{document}